# Measurements of Terahertz Generation in a Metallic, Corrugated Beam Pipe*

K.L.F. Bane, SLAC, Menlo Park, CA, USA
S. Antipov, ANL, Argonne, IL, USA
M. Fedurin, K. Kusche, C. Swinson, BNL, Upton, NY, USA
D. Xiang, Jiao Tong University, Shanghai, China

*Presented at the 7th International Particle Accelerator Conference, IPAC2016
Busan, Korea, 8–13 May 2016*

\* Work supported by Department of Energy contract DE–AC02–76SF00515 and by US DOE SBIR program grant No. DE-SC0009571.

# MEASUREMENTS OF TERAHERTZ GENERATION IN A METALLIC, CORRUGATED BEAM PIPE

K.L.F. Bane*, SLAC, Menlo Park, CA, USA
S. Antipov, ANL, Argonne, IL, USA
M. Fedurin, K. Kusche, C. Swinson, BNL, Upton, NY, USA
D. Xiang, Jiao Tong University, Shanghai, China

*Abstract*

A method for producing narrow-band THz radiation proposes passing an ultra-relativistic beam through a metallic pipe with small periodic corrugations. We present results of a measurement of such an arrangement at Brookhaven's Accelerator Test Facility (ATF). Our pipe was copper and was 5 cm long; the aperture was cylindrically symmetric, with a 1 mm (radius) bore and a corrugation depth (peak-to-peak) of 60 um. In the experiment we measured both the effect on the beam of the structure wakefield and the spectral properties of the radiation excited by the beam. We began by injecting a relatively long beam compared to the wavelength of the radiation to excite the structure, and then used a downstream spectrometer to infer the radiation wavelength. This was followed by injecting a shorter bunch, and then using an interferometer (also downstream of the corrugated pipe) to measure the spectrum of the induced THz radiation.

## INTRODUCTION

There is great interest in having a source of short, intense pulses of terahertz radiation. There are laser-based sources of such radiation, capable of generating few-cycle pulses with frequency over the range 0.5–6 THz and energy of up to 100 $\mu$J [1]. And there are beam-based sources, utilizing short, relativistic electron bunches. One beam-based method impinges an electron bunch on a thin metallic foil and generates coherent transition radiation (CTR). Recent tests of this method at the Linac Coherent Light Source (LCLS) have obtained single-cycle pulses of radiation that is broad-band, centered on 10 THz, and contains > 0.1 mJ of energy [2]. Another beam-based method generates narrow-band THz radiation by passing a bunch through a metallic pipe coated with a thin dielectric layer [3]- [5].

Another, similar method for producing narrow-band THz radiation has proposed passing the beam through a metallic pipe with small periodic corrugations [6], which is the subject of the present report. We consider here round geometry which will yield radially polarized THz (studies of this idea in flat geometry can also be found [7]). We present results of both measurements of the effect on the beam and the spectral properties of the radiation excited by the beam. We first use a relatively long beam—compared to the wavelength of the radiation—to excite the structure, and then use a downstream spectrometer to infer the wavelength of the THz. Then for a shorter bunch, by means of an interferometer also downstream of the corrugated pipe, we measure the spectrum of the induced THz radiation. Our experimental set-up was simple and not optimized for the efficient collection of the radiation (by *e.g.* the inclusion of tapered horns between the structure and the collecting mirror of the interferometer, as was done in Refs. [4], [5]). As such, the present experiment should be considered a proof-of-principle experiment for generating THz using a corrugated, metallic structure.

## TPIPE

Consider a metallic beam pipe with a round bore and small, rectangular (in longitudinal view) corrugations. The parameters are period $p$, (peak-to-peak) depth of corrugation $\delta$, corrugation gap $g$, and pipe radius $a$; where we consider small corrugations $(\delta, p) \ll a$ and also $\delta \gtrsim p$. Let us here assume $p = 2g$. It can be shown that a short, relativistic bunch, on passing through such a structure, will induce a wakefield that is composed of one dominant mode, of wave number $k \approx 2/\sqrt{a\delta}$, group velocity $v_g/c \approx 1 - 2\delta/a$, with $c$ the speed of light, and loss factor $\kappa \approx Z_0 c/(2\pi a^2)$, with $Z_0 = 377$ $\Omega$ [6]. In addition to the effect on the beam, a radiation pulse of the same frequency and of full length $\ell = 2\delta L/a$ ($L$ is pipe length) will follow the beam out the downstream end of the structure. One can see that, in order to generate a pulse of frequency $\sim$ 1 THz, both the bore radius and the corrugation dimensions must be small; with $a \sim 1$ mm, then $\delta \lesssim 10$ $\mu$m. For $a = 1$ mm, $\delta = 60$ $\mu$m, the pulse frequency $f \approx 0.4$ THz.

A corrugated structure, that we name TPIPE, was machined from two rectangular blocks of copper, each of dimension 2 cm by 1 cm by 5 cm on a side. Two 2-mm diameter cylindrical grooves were first machined in the long direction in each block. One groove in each block was meant to remain smooth, for the null test of the experiment. The other groove was further machined to yield corrugations with period 230 $\mu$m and peak-to-peak radius variation of 82.5 $\mu$m. Since rectangular corrugations of such small size were difficult to produce, the longitudinal profiles of the peaks and valleys were made circular with radius, respectively, of 41.3 $\mu$m and 90.7 $\mu$m (see Fig. 1, the lower right plot). Finally the two copper blocks were diffusion bonded together to yield one block with two cylindrically symmetric bores, one corrugated and one smooth.

We performed time-domain simulations with the Maxwell equation solving program ECHO [8], using a Gaussian bunch with $\sigma_z = 50$ $\mu$m, passing through the entire 5-cm-long structure with what we believe to be the actual geometry.

---

* kbane@slac.stanford.edu

The result is that a 6.5-mm-long THz pulse is generated, with frequency $f = 403$ GHz and quality factor $Q = 12$. For a 50 pC bunch, the average energy loss is 30 keV, 90% of which (or 1.2 $\mu$J) ends up in the THz pulse, and 10% in Joule heating of the structure walls.

## EXPERIMENT

TPIPE was tested with beam at the Accelerator Test Facility at Brookhaven National Laboratory. A schematic of the experimental layout is presented on Fig. 1. A 57 MeV electron beam was initially shaped using a mask [9] and then sent through TPIPE to generate a THz pulse. After leaving the structure, the radiation pulse diffracts and follows the electron beam, and further downstream some of it is reflected by an off-axis parabolic mirror into a Michelson interferometer for characterization. Note that the mirror is located 17.5 cm downstream of TPIPE, and that in front of the mirror is a 2.5 cm diameter iris, which limits the radiation that is collected. The intensity of the THz signal collected by the interferometer was measured by a LHe bolometer. Meanwhile, the electron beam passed through a 3 mm diameter hole in the mirror and then into the spectrometer for characterization.

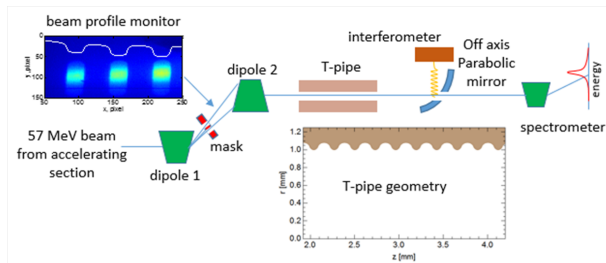

Figure 1: Sketch of the experimental layout. The profile of a section of TPIPE is given at the lower right.

To shape the beam we started by accelerating it off-crest in the accelerating section, in order to create a linear correlation between the longitudinal bunch coordinates $z$ and energy $E$. When the beam passes through a dipole magnet ("dipole 1" in Fig. 1) it becomes horizontally dispersed as in an energy spectrometer. A transverse mask is placed after the first dipole to block electrons of certain energies. A second dipole of opposing sign ("dipole 2") restores the beam to its original state, minus, however, the electrons that were blocked. However, due to the original $E$-$z$ correlation in the beam, the result is that the mask shape is left imprinted on the longitudinal charge distribution. In the experiment we monitored the shape of the beam immediately after the mask and were able to calibrate the image for the purpose of measuring the bunch length. This was done in the following way (with TPIPE kept retracted from the beam path): A mask with three periodic holes was inserted into the beam's path. The resulting three bunches excited a periodic signal by means of diffraction radiation on passing through the hole in the parabolic mirror. We measured the periodicity of the excited signal using the interferometer. Thus we were able to relate distances in pixels on the beam monitor image (after the mask) to longitudinal dimensions in microns on the shaped beam.

In addition, due to the $E$-$z$ correlation of the beam, the image in the downstream spectrometer also carries information about the beam's longitudinal shape. Therefore, distances on the spectrometer image can be related to longitudinal distances within the beam.

## RESULTS

If we pass a long bunch with a short rise time—both compared to TPIPE's wavelength—through the structure, the beam will become energy modulated at the structure frequency (see *e.g.* [10]). If the long bunch begins with an energy chirp, then the wavelength of the mode can be measured on the spectrometer screen. In Fig. 2 we compare the energy spectrum measurements of: the original beam, with TPIPE removed from its path (top), the beam passing through the corrugated pipe in TPIPE (middle), and the beam passing through the smooth tube (bottom). First we note that the results for the cases of no TPIPE and the smooth pipe are similar, showing a chirped streak with no modulation. With the corrugated pipe (the middle frame) we see a clear modulation. Based on the spacing of the energy modulation and the calibration, we estimate that the frequency of the TPIPE mode is $459 \pm 32$ GHz.

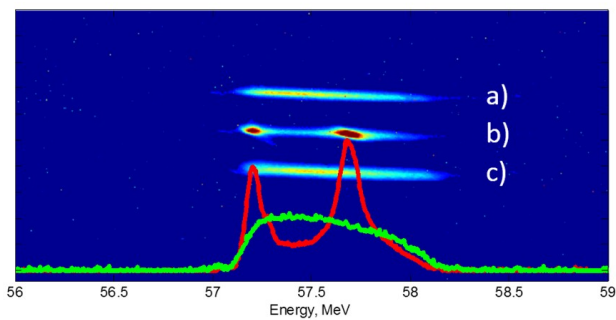

Figure 2: Spectrometer images: no structure (top), TPIPE (middle), smooth pipe (bottom). In all cases the beam is initially chirped, resulting in a horizontal streak on the screen. The green (red) line gives the energy distributions of a (b).

For the interferometer scans we need a shorter beam, generated using a mask, as described above. A fine, 20-mm-long interferometer scan, with the beam passing through TPIPE, is shown in Fig. 3a, where $ct$ gives the mirror movement distance. This plot is actually the composite of three scans. The bunch charge was $q = 50$ pC and the bunch length was measured, as discussed above, to have an rms value $\sigma_z = 50$ $\mu$m. The absolute value of the Fourier transform of this scan is given in Fig. 3b. We see a rather broad peak with a narrow horn on top. (Note that the dip at $f = 550$ GHz is found in all the scans we performed and is likely the absorption frequency of a gas in the environment.) The spectrum that is recorded has a low frequency cut-off given by the apertures in the interferometer, and a high frequency cut-off determined by the bunch spectrum cut-off in the diffraction radiation at the mirror hole. We choose the filter function [11]

$$\tilde{s}(k,\sigma,\zeta) = e^{-k^2\sigma^2}(1-e^{-k^2\zeta^2})^2, \quad (1)$$

with the two fitting parameters $\sigma$ and $\zeta$. The best fit, with $\sigma = 90$ $\mu$m and $\zeta = 134$ $\mu$m, is shown by the dashed curve in Fig. 3b.

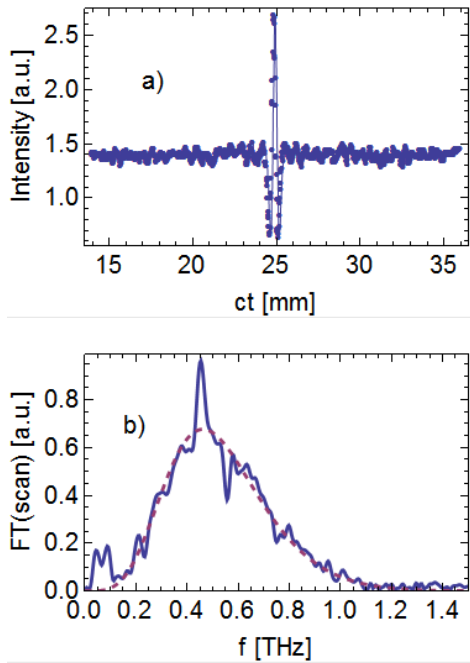

Figure 3: a) A fine, 20-mm interferometer scan of the radiation generated by TPIPE, and b) the absolute value of its Fourier transform. The dashed curve is discussed in the text.

If we consider this spectrum to be the sum of a broad and large diffraction radiation peak and a narrow peak representing TPIPE's resonance frequency, then we find that the THz signal has a central frequency of $f = 454$ GHz and a quality factor of $Q = 17$. The frequency agrees with the spectrometer measurement discussed above, but it does not agree with the ECHO calculations.

The interferometer results for the case of the beam passing through the smooth tube in TPIPE is given in Fig. 4 (the results for TPIPE removed from the path of the beam are similar). These scans were performed late in the data taking, when the helium was nearly depleted; thus there is less intensity in the signal. In addition, the scan only included a mirror movement of 8 mm. These two factors account for the measurement being noisier than the previous one. We again see the dip at $f = 550$ GHz, but this time there is no narrow-band horn or signal on top of the broad distribution.

There are aspects of the results of the measurements that, at the moment, are not understood. For example, why does the measured frequency disagree with the ECHO result by 10%? One possibility is that we don't accurately know the structure geometry. Another question is that the THz signal in the interferometer scan is smaller, compared to the background, than we expect. Such questions and more details of these measurements can be the subject of a future report.

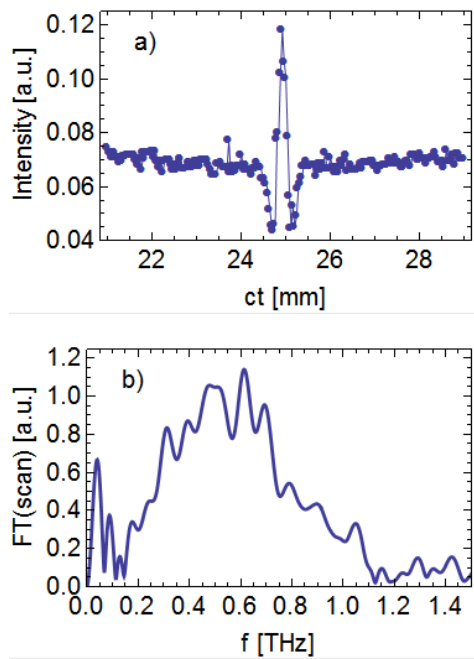

Figure 4: Interferometry scan and spectrum for the case of the beam passing through the smooth pipe of TPIPE.

## ACKNOWLEDGMENTS

We thank Makino Machine Tools for machining TPIPE for us free of charge, and G. Bowden, the engineer on the TPIPE project, for his careful work. We thank the Accelerator Test Facility staff at BNL for engineering support during the experiment. Euclid Beamlabs LLC acknowledges support from US DOE SBIR program grant No. DE-SC0009571. Work was partially supported by Department of Energy contract DE–AC02–76SF00515.